\documentclass[aps,twocolumn,pra,superscriptaddress,showpacs,tightenlines]{revtex4}
\usepackage{amssymb}
\usepackage{amsmath}
\usepackage{graphicx}
\usepackage{txfonts}

\begin{document}

\title{Asymmetric coherent transmission for single particle diode and
gyroscope}
\author{S. Yang}
\affiliation{Key Laboratory of Frontiers in Theoretical Physics,
Institute of Theoretical Physics,\\ Chinese Academy of Sciences,
Beijing 100190, China}
\author{Z. Song}
\affiliation{School of Physics, Nankai University, Tianjin 300071, China}
\author{C. P. Sun}
\affiliation{Key Laboratory of Frontiers in Theoretical Physics,
Institute of Theoretical Physics,\\ Chinese Academy of Sciences,
Beijing 100190, China}

\begin{abstract}
We study the single particle scattering process in a coherent multi-site
system consisting of a tight-binding ring threaded by an Aharonov-Bohm flux
and several attaching leads. The asymmetric behavior of scattering matrix is
discovered analytically in the framework of both Bethe Ansatz and Green's
function formalism. It is found that, under certain conditions, a three-site
electronic system can behave analogous to a perfect semiconductor diode
where current flows only in one direction. The general result is also valid
for a neutral particle system since the effective magnetic flux may be
implemented by a globe rotation. This observation means that the three-site
system can serve as an orientation measuring gyroscope due to the
approximate linear dependence of the current difference of two output leads
on the rotational angular velocity.
\end{abstract}

\pacs{73.23.-b, 05.60.Gg, 85.35.Ds}
\maketitle



\section{Introduction}

Single particle quantum devices work essentially in a quantum
mechanical way by using the whole features of quantum states,
especially the phases of quantum states. A typical example is the
single electron transistor (SET)
\cite{Averin,Kastner9293,Meirav95,Ashoori96,Beenakker} and its
photon analogue \cite{Lukin07,ZhouLan08}, which plays a central role
in quantum manipulation and quantum measurement. A SET is a
mesoscopic system that allows confined electrons to tunnel to the
metallic leads. It turns on and
off again every time one electron is added to the isolated region \cite%
{Kastner9293}. Unlike the conventional transistors, which can be understood
by using classical concepts, the SET is a quantum mechanical one in
substance. It can be utilized for measuring the quantum effects in Josephson
junction superconductor circuit and nano-mechanics resonators \cite%
{Shnirman9801,Schwab03,Korotkov05}.

In this article, we will pay attention to another single particle quantum
device, which can be understood as the quantum analogue of the conventional
diode device. In practice, diode devices are indispensable for building
various electronic circuits in the traditional electronics. Accordingly,
once a coherent diode device has been implemented, a single-particle quantum
circuit may be further realized in the level of single quantum state by
using of the two basic elements, single particle transistor and diode.
Compared with the traditional electrocircuits based on the density
distribution of electrons, the new quantum circuit makes full use of the
quantum properties including quantum phases in information processing. In
this sense, we can also regard it as a necessary element in quantum
information science and technology.

A major research effort in recent years is to seek a simplest single
particle system with diode features
\cite{Datta92,Datta02,Citro08,Pepino09}, which is characterized by
the asymmetric performance of transmission coefficients along the
opposite directions. In the present investigation, we find that a
three-site tight-binding ring system threaded by an Aharonov-Bohm
(AB) magnetic flux has analogy to a perfect diode in a wide range of
parameters.

In addition, our studies are not restricted to the charged particle case.
For a neutral particle system, an effective magnetic flux is induced by
applying a globe rotation \cite{Bhat06}, which can also break the time
reversal symmetry and lead to the asymmetric transmissions. It is shown that
the difference of the two output currents is linearly proportional to the
rotational angular velocity approximately, which behaves as a new kind of
extended Sagnac gyroscope \cite{Sagnac13,Post67}.

Decades ago, B\"{u}ttiker \textit{et al.} have found that the conductance of
a multi-terminal sample is asymmetric in the presence of an AB flux \cite%
{Buttiker86,Azbel81,Buttiker85}. Realization of such mechanism in
quantum device requires its size being smaller than the
phase-breaking length. Thus seeking a minimized system with
asymmetric feature is attractive in practice. A discrete system may
be a good candidate to accomplish such a task. It is of both
theoretical and practical importance to study this theory in a
discrete system. Moreover, we show practical applications of the
quantum interference effect as single electron diode and single
particle gyroscope. In the general discussion part, it is
interesting to find that some special configurations may protect the
transmission coefficients from symmetry breaking even if the AB
magnetic flux is imposed.

This paper is organized as follows. In Sec. II, the general $N$-site
ring-shaped model with attaching leads is presented, and an
analytical method for calculating the transmission coefficients is
introduced. In Sec. III, we study the asymmetric transmission
behavior of the $3$-site system by using the exact solution. In Sec.
IV, a single particle gyroscope and its physical realization in
optical lattice is proposed. In Sec. V, we give a general discussion
about the condition of asymmetric transmission by making use of
Green's function approach, which also confirms the results obtained
above. Conclusions are summarized at the end of the paper. In the
appendix, a proof is given to show the equivalence between the Bethe
Ansatz and the Green's function method.

\section{Model and its exact solution}
\begin{figure}[tbp]
\includegraphics[bb=41 522 550 777, width=8 cm, clip]{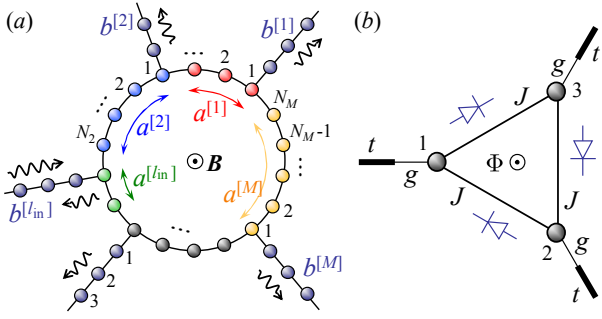}
\caption{(color online) (a) Configuration of the ring shaped
scattering system including $M$ arc chains $\{a^{[1]},a^{[2]},\dots
,a^{[M]}\}$ and $M$ attaching leads $\{b^{[1]},b^{[2]},\dots
,b^{[M]}\}$ threaded by a magnetic flux. The input and possible
output currents are marked by wavy arrows. (b) Schematic
illustration of the three-site chiral coherent scattering system
with diode features.} \label{model}
\end{figure}

The central system we concern is described by an $N$-site tight-binding ring
threaded by a magnetic flux shown in Fig. \ref{model}(a). The Hamiltonian
reads
\begin{equation}
H_{C}=-J\sum_{j=1}^{N}\left( e^{i\phi _{j}}a_{j}^{\dag }a_{j+1}+\mathrm{H.c.}%
\right) +\omega \sum_{j=1}^{N}a_{j}^{\dag }a_{j}.  \label{Hc}
\end{equation}%
Here, $a_{j}^{\dag }$ ($a_{j}$) is the fermion creation
(annihilation) operator at the $j$th site. The external magnetic
field does not exert force on the Bloch electrons, but makes hopping
integral $J$\ between sites $j$\ and $(j+1)$\ pick up an AB phase
factor $\exp \left( i\phi _{j}\right) $ in the Peierls approximation
\cite{Peierls33}, where
\begin{equation}
\phi _{j}=\frac{2\pi }{\phi _{0}}\int_{j}^{j+1}\mathbf{A}\cdot d\mathbf{l},
\end{equation}%
$\phi _{0}=hc/e$ is the flux quanta and $\Phi =\sum_{j=1}^{N}\phi _{j}$ is
the total magnetic flux; $\omega $ is the chemical potential of the central
system. In the same figure, $M$ half-infinite tight-binding leads are
attached to $M$ sites on the ring, and the corresponding Hamiltonian are%
\begin{equation}
H_{L}=\sum_{l=1}^{M}H_{l}=-\sum_{l=1}^{M}\left( ga_{j_{l}}^{\dag }b_{1}^{
\left[ l\right] }+t\sum_{i}b_{i}^{\dag \left[ l\right] }b_{i+1}^{\left[ l%
\right] }+\mathrm{H.c.}\right) .
\end{equation}%
Without loss of generality, we let $l=l_{in}$ lead as the input
lead, while all the other indexed $l\neq l_{in}$ leads are output
ones. The spin degree of freedom is omitted for notational brevity
here since the model does not contain interaction involving spin.

In order to investigate the scattering problem through such a system in the
framework of Bethe Ansatz \cite{ZhouLan08,JinLiang09,XiaJianBai92} more
efficiently, we regard the $N$-site ring as $M$ arc tight-binding chains
connected by the head and tail. As illustrated in Fig. \ref{model}(a), the
length of the $l$th arc chain $a^{\left[ l\right] }$ is $N_{l}$, and the $l$%
th lead $b^{\left[ l\right] }$ is joined with the $1$th site of the $l$th
arc chain. The total number of sites on the ring satisfies $%
\sum_{l=1}^{M}N_{l}=N$. Then the Hamiltonian $H=H_{C}+H_{L}$ is rearranged
with
\begin{eqnarray}
H_{C} &=&-J\sum_{l=1}^{M}\left[ \sum_{j=1}^{N_{l}-1}e^{i\phi
_{j}^{[l]}}a_{j}^{\dag \left[ l\right] }a_{j+1}^{\left[ l\right] }+e^{i\phi
_{N_{l-1}}^{[l-1]}}a_{N_{l-1}}^{\dag \left[ l-1\right] }a_{1}^{\left[ l%
\right] }+\mathrm{H.c.}\right] ,  \notag \\
H_{L} &=&-\sum_{l=1}^{M}\left( ga_{1}^{\dag \left[ l\right] }b_{1}^{\left[ l%
\right] }+t\sum_{i}b_{i}^{\dag \left[ l\right] }b_{i+1}^{\left[ l\right] }+%
\mathrm{H.c.}\right) .
\end{eqnarray}%
The scattering wave function is supposed to be%
\begin{eqnarray}
\psi _{a}^{\left[ l\right] }\left( j\right)  &=&A_{1}\left( l\right)
e^{i\left( qj-\sum_{m=1}^{j-1}\phi _{m}^{[l]}\right) }+A_{2}\left( l\right)
e^{-i\left( qj+\sum_{m=1}^{j-1}\phi _{m}^{[l]}\right) },  \notag \\
\psi _{b}^{\left[ l\right] }\left( j\right)  &=&B\left( l\right)
e^{ikj}+\delta _{l,l_{in}}e^{-ikj},
\end{eqnarray}%
where $B(l_{in})$ actually defines the reflection coefficient%
\begin{equation}
R_{l_{in}}=\left\vert B\left( l_{in}\right) \right\vert ^{2},
\end{equation}%
while $B\left( l\right) $ for $l\neq l_{in}$ \ give the transmission
coefficient from the $l_{in}$th lead to the $l$th lead,
\begin{equation}
T_{l_{in},l}=\left\vert B\left( l\right) \right\vert ^{2}
\end{equation}%
The coefficients $A_{1}\left( l\right) $, $A_{2}\left( l\right) $ and $%
B\left( l\right) $ should satisfy the connecting conditions and the Schr\"{o}%
dinger equation%
\begin{eqnarray}
&&\psi _{a}^{\left[ l-1\right] }\left( N_{l-1}+1\right) =\psi _{a}^{\left[ l%
\right] }\left( 1\right) ,  \notag \\
&&-g\psi _{b}^{\left[ l\right] }(1)-J\left[ \!e^{i\phi _{1}^{\left[ l\right]
}}\psi _{a}^{\left[ l\right] }(2)+e^{-i\phi _{N_{l-1}}^{\left[ l-1\right]
}}\psi _{a}^{\left[ l-1\right] }(N_{l-1})\right]   \notag \\
&&=(E-\omega )\psi _{a}^{\left[ l\right] }(1),  \notag \\
&&-t\psi _{b}^{\left[ l\right] }\left( 2\right) -g\psi _{a}^{\left[ l\right]
}\left( 1\right) =E\psi _{b}^{\left[ l\right] }\left( 1\right) .  \label{BA}
\end{eqnarray}%
The relation between $k$ and $q$ is also given by the Schr\"{o}dinger
equation,%
\begin{eqnarray}
&&\!-J\left[ e^{i\phi _{j}^{[l]}}\psi _{a}^{\left[ l\right] }\!\left(
j+1\right) +e^{-i\phi _{j-1}^{[l]}}\psi _{a}^{\left[ l\right] }\!\left(
j-1\right) \right] \!\!=\left( \!E-\omega \right) \psi _{a}^{\left[ l\right]
}\!\left( j\right) ,  \notag \\
&&-t\left[ \psi _{b}^{\left[ l\right] }\!\left( j+1\right) +\psi _{b}^{\left[
l\right] }\!\left( j-1\right) \right] \!=\!E\psi _{b}^{\left[ l\right]
}\!\left( j\right) ,
\end{eqnarray}%
for $j\neq 1$, i.e.,%
\begin{equation}
E=-2t\cos k=-2J\cos q+\omega .  \label{Ekq}
\end{equation}%
Then the $3M$ coefficients $\left\{ A_{1}\left( l\right) ,A_{2}\left(
l\right) ,B\left( l\right) \right\} $ are fully determined by solving the
above $3M$ independent equations in Eq. (\ref{BA}). It shows that the Bethe
Ansatz method can provide the exact wave function of the scattering state.%
\textbf{\ }For small size system, the explicit wave function allows us to
clarify the mechanism of the asymmetric transmission from the viewpoint of
interference.

\section{Asymmetric coherent transmission}

Hereafter, we mainly consider the simplest case of $N=3$ and $M=3$ shown in
Fig. \ref{model}(b). The Bethe Ansatz approach gives the exact solution of
transmission and reflection coefficients as

\begin{eqnarray}
&&T_{13}=T_{32}=T_{21}=T_{R},  \notag \\
&&T_{12}=T_{23}=T_{31}=T_{L},  \notag \\
&&R_{1}=R_{2}=R_{3}=1-T_{12}-T_{13}=R,  \label{TR}
\end{eqnarray}%
where%
\begin{eqnarray}
T_{R} &=&\frac{4g^{4}J^{2}t^{2}}{\Xi }\sin ^{2}k\left\vert Jte^{-i\Phi
}+\Theta \right\vert ^{2},  \notag \\
T_{L} &=&\frac{4g^{4}J^{2}t^{2}}{\Xi }\sin ^{2}k\left\vert Jte^{i\Phi
}+\Theta \right\vert ^{2},  \notag \\
\Theta &=&2t^{2}\cos k-g^{2}e^{ik}+t\omega ,  \notag \\
\Xi &=&\left\vert 2J^{3}t^{3}\cos \Phi +3J^{2}t^{2}\Theta -\Theta
^{3}\right\vert ^{2}.  \label{TR1}
\end{eqnarray}

We notice that in general, the transmission coefficients $T_{R}$ and $T_{L}$
from one lead to the other two are not identical, i.e., the current flow
across two arbitrary leads is unidirectional. On the other hand, the
clockwise coefficients $T_{13},T_{32},$and $T_{21}$, defined as $T_{R}$,
have the same value, so do the anti-clockwise coefficients $%
T_{12},T_{23},T_{31}$ defined as $T_{L}$. Namely, the transmission
coefficients have dextrorotary and levorotatory characteristics. We will
discuss the general asymmetric scattering problem in detail in Sec. V. We
will see that this kind of feature is reasonable for a non-bipartite system
with broken time-reversal symmetry induced by the external magnetic flux.
\begin{figure}[tbp]
\includegraphics[bb=45 282 550 792, width=8 cm, clip]{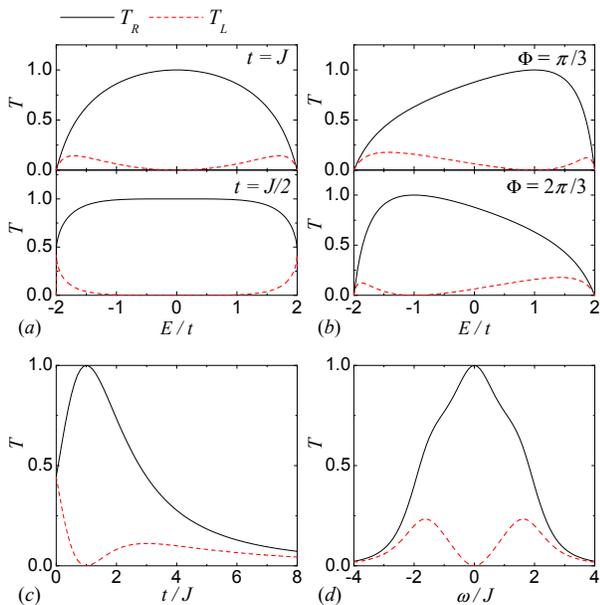}
\caption{(color online) Typical cases of transmission coefficients
in the clockwise direction $T_{R}$ (black solid lines) and
anti-clockwise direction $T_{L}$ (red dashed lines). (a) $T_{R}$ and
$T_{L}$ with respect to $E/t$ in the case of $t=J$ (upper panel) and
$t=J/2$ (lower panel), where $g^{2}/t=J$, $\Phi =\protect\pi /2$,
and $\protect\omega =0$. It shows that the system behaves like a
perfect diode at $E=0$ (upper panel) and within a wide range $ E\in
\lbrack -t,t]$ (lower panel). (b) Transmission spectrum in the case
of $ g=t=J$ and $\protect\omega =0$ with $\Phi =\protect\pi /3$
(upper panel) and $\Phi =2\protect\pi /3$ (lower panel). The perfect
diode energy can be adjusted by the magnetic flux $\Phi $. (c) Plots
of $T_{R}$ and $T_{L}$ at $ E=0$ as a function of $t/J$ with
parameters $g=t$, $\Phi =\protect\pi /2$ and $\protect\omega =0$.
The transmission coefficients of two directions coincide in the
limit of both $t\rightarrow 0$ and $t\rightarrow \infty $, but
diverge dramatically at the critical point $t=J$. (d) When
$g^{2}/t=J$, $ \Phi =\protect\pi /2$, and $E=0$, the diode feature
is optimal at the resonant point $\protect\omega /J=0$.}
\label{Transmission}
\end{figure}

Now we concentrate on some special cases of the $3$-site ring to exemplify
this feature.

(1) $g^{2}/t=J$, $\Phi =\pi /2$, and $\omega =0$, the Bloch electrons are
injected at $E=0$ with momentum $k=\pi /2$. Eq. (\ref{TR}) and (\ref{TR1})
becomes
\begin{equation}
T_{R}=1,T_{L}=0,R=0.
\end{equation}%
It means that the current can only flow in the clockwise direction with no
reflection. Thus the system is a perfect diode if we regard the $1$th lead
as the source and $2$th lead as the drain. The profiles of the corresponding
transmission spectra, $T_{R}$\ and $T_{L}$\ as functions of $E$\ for $t=J$\
and $t=J/2,$\ are plotted in Fig. \ref{Transmission}(a). One can see that,
when $t=J/2$ and $g^{2}/t=J$, the diode is perfect even if the input energy $%
E$ is shifted within the region $\left[ -t,t\right] $. This shows the great
tolerant as a perfect quantum device.

(2) $g=t=J$ and $\omega =0$, the input energy is determined by $k$ as $%
E=-2t\cos k$,
\begin{eqnarray}
T_{R} &=&1,T_{L}=0,R=0,\text{ for }k=\pi -\Phi +2n\pi ,  \notag \\
T_{R} &=&0,T_{L}=1,R=0,\text{ for }k=\Phi -\pi +2n\pi ,  \notag \\
T_{R} &=&0,T_{L}=0,R=1,\text{ for }k=n\pi .
\end{eqnarray}%
Therefore, for a given $E$, we can get a perfect diode device by tuning $%
\Phi $. The transmission spectrum of the $\Phi =\pi /3$ and $2\pi /3$ cases
are shown in \ref{Transmission}(b) to give a visual impression.

(3) $g=t$, $\Phi =\pi /2$ and $\omega =0$, we plot $T_{R}$ and $T_{L}$ at $%
E=0$ with respect to $t$ in Fig. \ref{Transmission}(c). The point of $t=J$
denotes the diode case. Beyond that critical point, the difference between $%
T_{R}$ and $T_{L}$ gets smaller. In the limit of $t\rightarrow 0$, $T_{R}$
and $T_{L}$ converged to $4/9$.

(4) $g^{2}/t=J$, $\Phi =\pi /2$, and $E=0$. The dependence of $T_{R}$ and $%
T_{L}$ on the frequency detuning $\omega $ is shown in Fig. \ref%
{Transmission}(d). It is apparent that the resonant case is advantageous to
forming a perfect diode.

A diode system based on the above mechanism is different from the classical
ones in semiconductor electronics. The conventional diode such as P-N
junction is made up based on the different density distribution of electrons
in the p-type and n-type materials. When the external voltage is absent and
the equilibrium is reached, there is a difference of chemical potentials
between the two materials. Consequently, current will flow readily in the
forward biased direction since the applied voltage decreases the barrier;
but not in the reverse biased direction because the barrier is raised. By
contrast, the single particle diode device presented here works in the
region with no chemical potential difference. It makes full use of the pure
quantum interference phenomenon induced by the Aharonov-Bohm effect.

\section{Single particle gyroscope}

So far we have seen that the threaded magnetic flux plays an important role
in controlling the amount and direction of currents. It can be applicable to
the more extended system. Actually, if the system is rotated, an effective
magnetic field will be induced in the rotating frame of references.
Therefore, the difference between the currents of the two output leads can
reflect the rotational angular velocity of the system. That is the basic
idea of making up a gyroscope by using the above mentioned asymmetric
scattering system. The analysis below can be applied to a neutral-particle
system by simply choosing $\phi =0$.

\begin{figure}[tbp]
\includegraphics[bb=43 504 542 770, width=8 cm, clip]{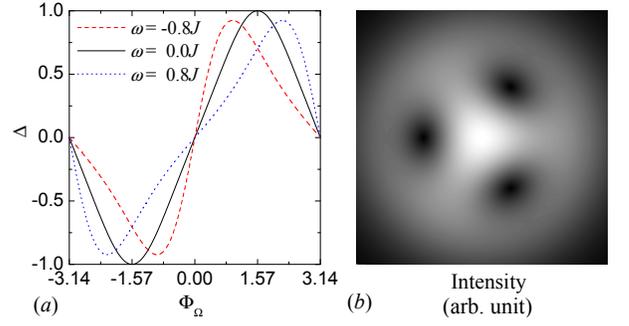}
\caption{(color online) (a) The difference of the transmission coefficients
on two directions $\Delta $ depends linearly on the total effective magnetic
flux $\Phi _{\Omega }$ around $\Phi _{\Omega }=0$ approximately. Here, $%
g^{2}/t=J$, $E=0$, the frequency detuning $\protect\omega $ is chosen as $%
\protect\omega =-0.8J$ (red dashed line), $\protect\omega =0$ (black
solid line), and $\protect\omega =0.8J$ (blue dot line). (b)
Intensity distribution of the three-site ring-shaped optical lattice
generated by two Laguerre-Gauss laser beams.} \label{gyroscope}
\end{figure}

For a rotating system with angular frequency $\Omega $, an additional term

\begin{equation}
H_{R}=-\Omega L_{z}=-\Omega K\sum_{j=1}^{N}\left( ie^{i\phi _{j}}a_{j}^{\dag
}a_{j+1}+\mathrm{H.c.}\right) ,
\end{equation}%
should be added on the Hamiltonian \cite{Bhat06} in the non-inertial frame,
where $K$ is a constant depends on the geometry of the central system. In
the meantime, the Hamiltonian of leads $H_{L}$ remains the same since the
leads are along the radial directions. Then the Hamiltonian of the central
system becomes $H_{C}^{\prime }=H_{C}+H_{R}$ or%
\begin{equation}
H_{C}^{\prime }=-J_{\Omega }\sum_{j=1}^{N}\left[ e^{i\left( \phi _{j}+\phi
_{\Omega }\right) }a_{j}^{\dag }a_{j+1}+\mathrm{H.c.}\right] ,
\end{equation}%
where $J_{\Omega }=\sqrt{J^{2}+\Omega ^{2}K^{2}}$, $\tan \phi _{\Omega
}=\Omega K/J$. The total effective magnetic flux
\begin{equation}
\Phi _{\Omega }=\sum_{j=1}^{3}\phi _{j}+3\phi _{\Omega }=\Phi +3\phi
_{\Omega }
\end{equation}%
depends on the angular frequency $\Omega $. For simplicity, we only focus on
the case of $E=0$ injection. If the total effective magnetic flux is absent,
i.e., $\Phi _{\Omega }=0$, the transmission and reflection are symmetric due
to the fact $T_{R}=T_{L}=T(\omega )$ with
\begin{equation}
T(\omega )=\frac{4g^{4}J^{2}t^{2}}{\left[ g^{4}+t^{2}\left( \omega
-2J\right) ^{2}\right] \left[ g^{4}+t^{4}\left( \omega +J\right) ^{2}\right]
}.
\end{equation}%
Then we have $\Delta =T_{R}-T_{L}=0$, which characterizes the asymmetrical
feature. For small effective magnetic flux $\Phi _{\Omega }\approx 0$, we
have%
\begin{equation}
\Delta =\frac{4Jtg^{2}T(\omega )}{g^{4}+t^{4}\left( \omega +J\right) ^{2}}%
\Phi _{\Omega }+O\left( \Phi _{\Omega }\right) ^{3},
\end{equation}%
i.e., the current difference $\Delta $\ is a linear function of $\Phi
_{\Omega }$\ in the vicinity of $\Phi _{\Omega }=0$. For a typical case $g=%
\sqrt{Jt}$, we plot $\Delta $\ as a function of $\Phi _{\Omega }$\ in Fig. %
\ref{gyroscope}(a). In the resonant case $\omega =0$, it shows that this
gyroscope have the advantage of good linear response within a wide range of $%
\Phi _{\Omega }$. In practice, in the case of extremely large
rotation frequency $\Omega $, a compensate magnetic flux can be
added to ensure it works in the linear region. For some off-resonant
cases, e.g., $\omega =-0.8J $, the ratio of $\Delta $\ to $\Phi
_{\Omega }$\ at $\Phi _{\Omega }=0$\ is larger than the one of the
resonant case. The larger ratio implies a higher sensitivity, so it
is good for measurement.

As a result, the instantaneous value $\Omega \left( t\right) $ can be
obtained by measuring $\Delta \left( t\right) $ directly. The cumulative
rotation angle is just an integration of the angular velocity over time,%
\begin{equation}
\theta \left( \tau \right) =2\pi \int_{0}^{\tau }\Omega \left( t\right) dt.
\end{equation}%
Therefore, such a system can measure the angular velocity and orientation
precisely as a gyroscope.

Actually, the most famous modern gyroscope is the laser gyroscope based on
Sagnac effect \cite{Sagnac13,Post67} (Two waves propagating along the
opposite directions though a closed rotating ring will induce a relative
phase difference, and the position of the interference fringes depends on
the angular velocity). Our proposal is an extention of the Sagnac gyroscope
since the interference effect is also used. However, the difference of the
two output currents is measured here instead of observing the interference
pattern directly, which may enhance the sensitivity of the results.

Our asymmetric coherent scattering system may be realized by trapping atoms
in a ring-shaped optical lattice. The optical lattices with periodic
boundary condition have been produced by the superposition of two
Laguerre-Gauss (LG) laser modes theoretically and experimentally \cite%
{Amico05,Arnold07}. The LG beam with frequency $\omega $, wave vector $k$,
and amplitude $E_{0}$ propagating along the $z$ axis has the form%
\begin{eqnarray}
LG_{l}^{p}\left( \omega \right) &=&E_{0}f_{pl}\left( r\right) e^{i\phi
l}e^{i\left( \omega t-kz\right) },  \notag \\
f_{pl}\left( r\right) &=&\left( -1\right) ^{p}\sqrt{\frac{2p!}{\pi \left(
p+\left\vert l\right\vert !\right) }}x^{\left\vert l\right\vert
}L_{p}^{l}\left( x^{2}\right) e^{-x^{2}/2},
\end{eqnarray}%
where $x=\sqrt{2}r/r_{0}$ and $r_{0}$ is the beam waist. $L_{p}^{l}(\cdot )$
is the generalized Laguerre polynomial. To generate a 1D optical lattice
ring with $N$ traps, parameters are chosen as $p_{1}=p_{2}=0$, $l_{1}=0$, $%
l_{2}=N$, and $x_{1}=1.3x_{2}$. The intensity distribution of the $N=3$ case
is shown in Fig. \ref{gyroscope}(b).

\section{Green's function approach and general discussion for asymmetric
transmission}

In this section, we generally consider in which case the asymmetric
transmission can happen for a general multi-terminal tight-binding system.
For an arbitrary single-particle tight-binding model with attaching leads,
we will derive the condition of the asymmetric transmission $T_{pq}\neq
T_{qp}$\ between lead $p$\ and $q$\ by using the Green's function formalism
introduced in Ref. \cite{Datta} and references therein.

We first briefly review Green's function method for multi-terminal
system. The retarded Green's function of the system as a function of
input
energy $E=-2t\cos k$ is%
\begin{equation}
G^{R}=\frac{1}{E-H_{C}-\Sigma _{leads}},  \label{GR}
\end{equation}%
where $H_{C}$ is the Hamiltonian of the central system, and $\Sigma _{leads}$
as the total self-energy denotes the contribution of the half-infinite
leads, $\Sigma _{leads}=\sum_{p\in \{\text{lead}\}}\Sigma _{p}$. On the
single particle basis $\{a_{1}^{\dag }\left\vert 0\right\rangle $, $%
a_{2}^{\dag }\left\vert 0\right\rangle $, $\cdots $, $a_{N}^{\dag
}\left\vert 0\right\rangle \}$, both $H_{C}$ and $\Sigma _{p}$ are $N\times N
$ matrix. Especially, each $\Sigma _{p}$ has only one non-zero matrix
element at the cross point of row $p$ and column $p$,
\begin{equation}
\left( \Sigma _{p}\right) _{p,p}=-\frac{g^{2}}{t}e^{ik\left( E\right)
}=\Sigma _{0},  \label{selfenergy}
\end{equation}%
which is because the $p$th lead is only attached to the $p$th site of the
central system via $(-ga_{p}^{\dag }b_{1}^{\left[ p\right] }+\mathrm{H.c.})$%
. Then the advanced Green's function $G^{A}=G^{R\dag }$ and $\Gamma _{p}$
matrix $\Gamma _{p}=i\left[ \Sigma _{p}-\Sigma _{p}^{\dag }\right] $ are
obtained sequentially. The transmission spectrum from lead $p$ to lead $q$ ($%
q\neq p$) is given by
\begin{equation}
T_{pq}=\mathrm{Tr}\left[ \Gamma _{q}G^{R}\Gamma _{p}G^{A}\right] .
\label{Greenfunction}
\end{equation}%
Applying it to the ring system mentioned in Sec. II-IV, one can see that the
Green's function formalism gives the same exact solution as the one form
Bethe Ansatz method. The equivalence of the two approaches is proved in the
appendix.

In general, the Green's functions $G^{R}$ (Eq. (\ref{GR})) and $G^{R\dag }$
are complex. From Eq. (\ref{Greenfunction}), the transmission coefficient $%
T_{pq}$ is
\begin{eqnarray}
T_{pq} &=&\mathrm{Tr}\left[ \Gamma _{q}\left( \mathrm{Re}G^{R}+i\mathrm{Im}%
G^{R}\right) \Gamma _{p}\left( \mathrm{Re}G^{R\dag }+i\mathrm{Im}G^{R\dag
}\right) \right]   \notag \\
&=&(\Gamma _{q})_{qq}\mathrm{Re}(G^{R})_{qp}(\Gamma _{p})_{pp}\mathrm{Re}%
(G^{R\dag })_{pq}  \notag \\
&&-(\Gamma _{q})_{qq}\mathrm{Im}(G^{R})_{qp}(\Gamma _{p})_{pp}\mathrm{Im}%
(G^{R\dag })_{pq}  \notag \\
&=&(\Gamma _{p})_{pp}(\Gamma _{q})_{qq}\left[ \mathrm{Re}(G^{R})_{qp}^{2}+%
\mathrm{Im}(G^{R})_{qp}^{2}\right]   \notag \\
&=&(\Gamma _{p})_{pp}(\Gamma _{q})_{qq}\left\vert (G^{R})_{qp}\right\vert
^{2},  \label{Tpq}
\end{eqnarray}%
where we have use the identity
\begin{eqnarray}
\mathrm{Re}(G^{R})_{qp} &=&\mathrm{Re}(G^{R\dag })_{pq},  \notag \\
\mathrm{Im}(G^{R})_{qp} &=&-\mathrm{Im}(G^{R\dag })_{pq},
\end{eqnarray}%
together with the property that the $\Gamma $ matrix has only one non-zero
matrix element. Similarly, for the inverse transport,%
\begin{equation}
T_{qp}=(\Gamma _{p})_{pp}(\Gamma _{q})_{qq}\left\vert
(G^{R})_{pq}\right\vert ^{2}.
\end{equation}%
Therefore, once
\begin{equation}
\left\vert (G^{R})_{qp}\right\vert =\left\vert (G^{R})_{pq}\right\vert ,
\label{condition}
\end{equation}%
we have a symmetric scattering $T_{pq}=T_{qp}$. When this condition is not
satisfied, i.e., $\left\vert (G^{R})_{qp}\right\vert \neq \left\vert
(G^{R})_{pq}\right\vert $, an asymmetric scattering is obtained. Eq. (\ref%
{GR}) tells us that the condition is not only related to the intrinsic
symmetry of the central Hamiltonian $H$, but also depends on the features of
the leads. For example, if we only attach two leads on the same $3$-site
central ring investigated in Sec. III, the asymmetric transmission phenomena
will disappear. This result is in agreement with the viewpoint in Ref. \cite%
{Buttiker85} that the simplest example must involve at least three channels.

Since $H_{C}$ is Hermitian and $\Sigma _{leads}$ is diagonal, the term%
\begin{equation}
H_{eff}=E-H_{C}-\Sigma _{leads}
\end{equation}%
can be written as a matrix as%
\begin{equation}
H_{eff}=\left(
\begin{array}{ccccc}
c_{1} & d_{12} & d_{13} & \cdots  & d_{1N} \\
d_{12}^{\ast } & c_{2} & d_{23} & \cdots  & d_{2N} \\
d_{13}^{\ast } & d_{23}^{\ast } & \ddots  &  & \vdots  \\
\vdots  & \vdots  &  & c_{N-1} & d_{N-1,N} \\
d_{1N}^{\ast } & d_{2N}^{\ast } & \cdots  & d_{N-1,N}^{\ast } & c_{N}%
\end{array}%
\right) .
\end{equation}%
The diagonal matrix elements are noted by $c$, while the off-diagonal ones
are denoted by $d$. The inverse matrix of $H_{eff}$ has the form
\begin{equation}
G^{R}=\frac{D}{\mathrm{Det}(H_{eff})},
\end{equation}%
where $D=\mathrm{Adj}(H_{eff})$ is the adjugate matrix of $H_{eff}$, which
is the transpose of the matrix of cofactors. Since the determinant$\mathrm{\
Det}(H_{eff})$ is a constant, our problem is reduced to the condition of the
identity $\left\vert D_{qp}\right\vert =\left\vert D_{pq}\right\vert $.
Since a matrix and its transpose have the same determinant, we notice that $%
D_{qp}$\ and $D_{pq}$\ have the form%
\begin{eqnarray}
D_{qp}\!\! &=&\!F\left( c_{1},\cdots \!,c_{N};d_{12},\cdots\!
,d_{N-1,N};d_{12}^{\ast
},\cdots \!,d_{N-1,N}^{\ast }\right) , \notag\\
D_{pq}\!\!&=&\!F\left( c_{1},\cdots \!,c_{N};d_{12}^{\ast },\cdots\!
,d_{N-1,N}^{\ast };d_{12},\cdots \!,d_{N-1,N}\right) .
\end{eqnarray}%
Here, $F$ is a polynomial function constructed by $c$ and $d$. Generally
speaking, if all of the $c$ and $d$ are non-zero random complex numbers, $%
\left\vert D_{qp}\right\vert $ is probably not equal to $\left\vert
D_{pq}\right\vert $. However, under some specific conditions, the asymmetric
transmission will disappear. Now we\ focus on the resonant systems (there is
no chemical potential difference between the leads and the central lattice)
and take some typical cases as an illustration.

(1) Obviously, when there is no external magnetic flux breaks the time
reversal symmetry, we have all the $d=d^{\ast }$, then $D_{qp}=D_{pq}$, the
condition Eq. (\ref{condition}) is automatically satisfied.

(2) When all the $c$ are real, i.e., $c=c^{\ast }$, $D_{qp}=D_{pq}^{\ast }$,
the condition Eq. (\ref{condition}) is also satisfied. Because of Eq. (\ref%
{selfenergy}) and (\ref{Ekq}), this case requires that the incident momentum
$k$ is either $0$ or $\pi $, which corresponds to injecting an electron at
the top or bottom of the Bloch energy band. Thus it is not very easy to be
realized in practice.

(3) A realizable but not trivial case: The symmetric scattering can be
obtained even though the magnetic field is present if the geometry of the
lattice is bipartite and electrons are injected at $E=0$. A bipartite
lattice is a lattice that can be divided into A and B sublattices, such that
a site in A is connected only with sites in B and vice versa. At this time,
the function $F$ is a polynomial contains only either even order of $c$ or
odd order of $c$. In other words, if one term of $F$ is a product of even
number of $c$ and some $d$, no term of $F$ contains the product of odd
number of $c$, thus $F$ is called even with respect to $c$. The situation is
similar when $F$ is odd with respect to $c$. The momentum for $E=0$ is $%
k=\pi /2$, then all the $c$ are pure imaginary, $c=-c^{\ast }$. Therefore, $%
D_{qp}=\left( -1\right) ^{\mathrm{even}}D_{pq}^{\ast }$ if $F$ is even with
respect to $c$, while $D_{qp}=\left( -1\right) ^{\mathrm{odd}}D_{pq}^{\ast }$
if $F$ is odd with respect to $c$. For both cases, we have $\left\vert
(G^{R})_{qp}\right\vert =\left\vert (G^{R})_{pq}\right\vert $ which
represents a symmetric scattering.

(4) For the bipartite central lattice, $T_{qp}$ equals to $T_{pq}$ even
though the input energy $E\neq 0$ if one of the following conditions is
satisfied. (i) $N$ is odd, there is no lead connected to the sublattice with
$\left( N-1\right) /2$ sites, or there are no more than two leads connected
to the sublattice with $\left( N+1\right) /2$ sites. (ii) $N$ is even, there
are no more than one lead connected to either sublattice A or sublattice B.

\section{Conclusion}

In this paper, we take the three-site tight-binding ring as an
explicit example to study the scattering problem of some coherent
nanostructure with or without time reversal symmetry by virtue of
exact solutions. It is discovered that, induced by the threaded or
effective magnetic flux, the asymmetric transmission can happen in
such a system. With this interesting property, on the one hand, such
a system behaves as a quantum diode for a charged particle due to
the pure quantum interference effect. On the other hand, it can also
be served as a gyroscope for either a charged or neutral particle
when a global rotation providing an effective magnetic flux. We have
shown that the difference of the output currents is linear
proportional to the rotational angular velocity approximately. The
observable effects presented in this paper are hopeful to be
realized experimentally in the quantum dot system and the ultra-cold
atom optical lattice system. We also try to generalize the results
from the three-site system to the general ones. According to the
Green's function approach \cite{Datta}, it is demonstrated that the
intrinsic symmetries of the whole system, including the geometry of
lattice and leads, determine the symmetry of scattering coefficients
together. Protected by the bipartite configuration, a symmetry
scattering may be obtained even if the time reversal symmetry is
broken for the central system.

\acknowledgments

This work is supported by NSFC No. 10474104, 60433050, 10874091 and
No. 10704023, NFRPC No. 2006CB921205 and 2005CB724508.

\appendix*
\section{Equivalence between Bethe Ansatz method and Green's function method}

In this appendix, we prove that the Bethe Ansatz method and Green's function
method are equivalent for determining the transmission coefficients.

The Hamiltonian of a general central system with $N$ sites reads%
\begin{equation}
H_{C}=\sum_{j,j^{\prime }=1}^{N}\left( H_{C}\right) _{j,j^{\prime
}}a_{j}^{\dag }a_{j^{\prime }}.
\end{equation}%
We suppose that there is a set of $j$ defined as $\{$lead$\}=\left\{ j_{1}%
\text{, }j_{2}\text{, }\cdots \text{, }j_{M}\right\} $, where $2\leq M\leq N$%
. If $j\in \{$lead$\}$, a half infinite tight-binding lead is attached to
the $j$th site. The total Hamiltonian of leads is%
\begin{equation}
H_{L}=-\sum_{j\in \{\text{lead}\}}\left( ga_{j}^{\dag
}b_{1}^{[j]}+t\sum_{j^{\prime }=1}^{\infty }b_{j^{\prime }}^{\dag \lbrack
j]}b_{j^{\prime }+1}^{[j]}+\mathrm{H.c.}\right) .
\end{equation}%
According to the Bethe Ansatz method, the scattering state is supposed to be%
\begin{equation}
\left\vert \psi \right\rangle =\sum_{j}\psi _{a}\left( j\right) a_{j}^{\dag
}\left\vert 0\right\rangle +\sum_{j\in \{\text{lead}\}}\sum_{j^{\prime
}}\psi _{b}^{[j]}\left( j^{\prime }\right) b_{j^{\prime }}^{\dag \lbrack
j]}\left\vert 0\right\rangle ,
\end{equation}%
where
\begin{equation}
\psi _{b}^{[j]}\left( j^{\prime }\right) =B\left( j\right) e^{ikj^{\prime
}}+\delta _{j,p}e^{-ikj^{\prime }}.  \label{Awaveb}
\end{equation}%
The $p$th lead has been chosen as the input lead. From the Schr\"{o}dinger
equation%
\begin{equation}
\left( H_{C}+H_{L}\right) \left\vert \psi \right\rangle =E\left\vert \psi
\right\rangle ,
\end{equation}%
$\psi _{a}\left( j\right) $ and $\psi _{b}^{[j]}\left( j\right) $ should
satisfy%
\begin{equation}
\sum_{j^{\prime }}\left( H_{C}\right) _{j,j^{\prime }}\psi _{a}\left(
j^{\prime }\right) -g\delta _{j,\{\text{lead}\}}\psi _{b}^{[j]}\left(
1\right) =E\psi _{a}\left( j\right)  \label{Ascheq}
\end{equation}%
for any $j$, and
\begin{eqnarray}
-g\psi _{a}\left( j\right) -t\psi _{b}^{[j]}\left( 2\right) &=&E\psi
_{b}^{[j]}\left( 1\right) ,  \notag \\
-t\left[ \psi _{b}^{[j]}\left( j^{\prime }+1\right) +\psi _{b}^{[j]}\left(
j^{\prime }-1\right) \right] &=&E\psi _{b}^{[j]}\left( j^{\prime }\right)
\label{Apsib}
\end{eqnarray}%
for $j\in \{$lead$\}$. Eq. (\ref{Awaveb}) and (\ref{Apsib}) give the
dispersion relation%
\begin{equation}
E=-2t\cos k,
\end{equation}%
together with the equation for $j\in \{$lead$\}$,
\begin{equation}
\frac{g}{t}\psi _{a}\left( j\right) =B\left( j\right) +\delta _{j,p}.
\end{equation}%
For future convenience, we define a column vector $X$,
\begin{eqnarray}
X &=&\left(
\begin{array}{cccc}
X_{1} & X_{2} & \cdots & X_{N}%
\end{array}%
\right) ^{\mathrm{T}},  \notag \\
X_{j} &=&\left( 1-\delta _{j,\{\text{lead}\}}\right) \left( \frac{g}{t}\psi
_{a}\left( j\right) \right) +\delta _{j,\{\text{lead}\}}B\left( j\right) .
\end{eqnarray}%
Eq. (\ref{Ascheq}) can be simplified as%
\begin{eqnarray}
&&\sum_{j^{\prime }}\left[ E\delta _{j,j^{\prime }}-\left( H_{C}\right)
_{j,j^{\prime }}-\left( -\frac{g^{2}}{t}e^{ik}\delta _{j,j^{\prime }}\delta
_{j,\{\text{lead}\}}\right) \right] X_{j^{\prime }}  \notag \\
&=&\left( H_{C}\right) _{j,p}-\delta _{j,p}\left( E+\frac{g^{2}}{t}%
e^{-ik}\right) ,  \label{Alineqs}
\end{eqnarray}%
which is a system of linear equations in the variables $X_{j}$. In the
matrix form, Eq. (\ref{Alineqs}) becomes%
\begin{equation}
H_{eff}X=W,
\end{equation}%
where%
\begin{equation}
H_{eff}=E-H-\Sigma _{leads}.
\end{equation}%
The matrix elements of $H_{eff}$, $W$, and $\Sigma _{leads}$ are
\begin{eqnarray}
\left( H_{eff}\right) _{j,j^{\prime }} &=&E\delta _{j,j^{\prime }}-\left(
H_{C}\right) _{j,j^{\prime }}-\left( -\frac{g^{2}}{t}e^{ik}\delta
_{j,j^{\prime }}\delta _{j,\{\text{lead}\}}\right) ,  \notag \\
W_{j} &=&\left( H_{C}\right) _{j,p}-\delta _{j,p}\left( E+\frac{g^{2}}{t}%
e^{-ik}\right) ,  \notag \\
\left( \Sigma _{leads}\right) _{j,j^{\prime }} &=&-\frac{g^{2}}{t}%
e^{ik}\delta _{j,j^{\prime }}\delta _{j,\{\text{lead}\}}=\Sigma _{0}\delta
_{j,j^{\prime }}\delta _{j,\{\text{lead}\}},
\end{eqnarray}%
where $\Sigma _{0}=-\left( g^{2}/t\right) e^{ik}$. The solution of
$X_{j}$ can be given by the Cramer's rule \cite{Cramer}. Without
loss of generality, we
focus on the solution of $X_{q}$ with $q\in \{$lead$\}$ but $q\neq p$,%
\begin{equation}
X_{q}=B\left( q\right) =\frac{\mathrm{Det}\left( H_{eff}^{\left( q\right)
}\right) }{\mathrm{Det}\left( H_{eff}\right) }.
\end{equation}%
Here, the matrix elements of $H_{eff}^{\left( q\right) }$ are%
\begin{equation}
\left( H_{eff}^{\left( q\right) }\right) _{j,j^{\prime }}=\left( 1-\delta
_{j^{\prime },q}\right) \left( H_{eff}\right) _{j,j^{\prime }}+\delta
_{j^{\prime },q}W_{j}.
\end{equation}%
The determinant of $H_{eff}^{\left( q\right) }$ equals to the one of another
matrix $H_{eff}^{\left( q\right) \prime }$, where%
\begin{eqnarray}
\left( H_{eff}^{\left( q\right) \prime }\right) _{j,j^{\prime }} &=&\left(
1-\delta _{j^{\prime },q}\right) \left( H_{eff}\right) _{j,j^{\prime
}}+\delta _{j^{\prime },q}W_{j}+\delta _{j^{\prime },q}\left( H_{eff}\right)
_{j,p}  \notag \\
&=&\left( 1-\delta _{j^{\prime },q}\right) \left( H_{eff}\right)
_{j,j^{\prime }}+i\delta _{j,p}\delta _{j^{\prime },q}\Gamma _{0},
\end{eqnarray}%
and
\begin{equation}
\Gamma _{0}=i\left( \Sigma _{0}-\Sigma _{0}^{\ast }\right)
\end{equation}%
is a real number. So that%
\begin{equation}
\mathrm{Det}\left( H_{eff}^{\left( q\right) }\right) =\mathrm{Det}\left(
H_{eff}^{\left( q\right) \prime }\right) =i\Gamma _{0}\Lambda _{pq}.
\end{equation}%
$\Lambda _{pq}$ is the $\left( p,q\right) $th algebraic cofactor of $H_{eff}$%
, which is actually $\left( -1\right) ^{p+q}$ times the determinant of the
submatrix obtained by removing $p$th row and $q$th column from $H_{eff}$.
Therefore, the transmission coefficient from lead $p$ to lead $q$ is%
\begin{equation}
T_{pq}=\left\vert B\left( q\right) \right\vert ^{2}=\left\vert \frac{i\Gamma
_{0}\Lambda _{pq}}{\mathrm{Det}\left( H_{eff}\right) }\right\vert
^{2}=\Gamma _{0}^{2}\left\vert \frac{\Lambda _{pq}}{\mathrm{Det}\left(
H_{eff}\right) }\right\vert ^{2}.  \label{ATpq}
\end{equation}%
The definition of the inverse of a matrix tells us that Eq. (\ref{Ascheq})
is actually%
\begin{equation}
T_{pq}=\Gamma _{0}^{2}\left\vert \left( H_{eff}^{-1}\right) _{qp}\right\vert
^{2}=\Gamma _{0}^{2}\left\vert \left( G^{R}\right) _{qp}\right\vert ^{2}.
\label{ATpqB}
\end{equation}

On the other hand, from Green's function method Eq. (\ref{Tpq}),
\begin{equation}
T_{pq}=(\Gamma _{p})_{pp}(\Gamma _{q})_{qq}\left\vert
(G^{R})_{qp}\right\vert ^{2},  \label{ATpqG}
\end{equation}%
where $(\Gamma _{p})_{pp}=(\Gamma _{q})_{qq}=\Gamma _{0}$. Then Eq. (\ref%
{ATpqG}) and (\ref{ATpqB}) are same, the Bethe Ansatz method is
equivalent to the Green's function method for determining the
transmission coefficients. Moreover, the Bethe Ansatz method can
also gives the exact wave function.


\begin{thebibliography}{99}
\bibitem{Averin} D. V. Averin and K. K Likharev, in \textit{Mesoscopic
Phenomena in Solids}, edited by B. L. Altshuler, P. A. Lee, and R. A. Webb
(Elsevier, Amsterdam, 1991).

\bibitem{Kastner9293} M. A. Kastner, Rev. Mod. Phys. \textbf{64}, 853
(1992); Phys. Today \textbf{46}, 24 (1993).

\bibitem{Meirav95} U. Meirav and E. B. Foxman, Semicond. Sci. Technol.
\textbf{10}, 255 (1995).

\bibitem{Ashoori96} R. C. Ashoori, Nature \textbf{379}, 413 (1996).

\bibitem{Beenakker} C. W. J. Beenakker, H. V. Houten, and A. A. M. Staring,
in \textit{Single Charge Tunneling}, edited by H. Grabert and M. H. Devoret
(NATO ASI Series B, Plenum, New York, 1991).


\bibitem{Lukin07} D. E. Chang, A. S. S{\o }rensen, E. A. Demler, and M. D.
Lukin, Nat. Phys. \textbf{3}, 807 (2007).

\bibitem{ZhouLan08} L. Zhou, Z. R. Gong, Y. X. Liu, C. P. Sun, and F. Nori,
Phys. Rev. Lett. \textbf{101}, 100501 (2008).


\bibitem{Shnirman9801} A. Shnirman and G. Sch{\o }n, Phys. Rev. B \textbf{57}%
, 15400 (1998); Y. Makhlin, G. Sch{\o }n, and A. Shnirman, Rev. Mod. Phys.
\textbf{73}, 357 (2001).

\bibitem{Schwab03} A. Hopkins, K. Jacobs, S. Habib, and K. Schwab, Phys.
Rev. B \textbf{68}, 235328 (2003).

\bibitem{Korotkov05} R. Ruskov, K. Schwab, and A. N. Korotkov, Phys. Rev. B
\textbf{71}, 235407 (2005).


\bibitem{Datta92} R. Lake and S. Datta, Phys. Rev. B \textbf{45}, 6670
(1992).

\bibitem{Datta02} T. Koga, J. Nitta, H. Takayanagi, and S. Datta, Phys. Rev.
Lett. \textbf{88}, 126601 (2002).

\bibitem{Citro08} R. Citro and F. Romeo, Phys. Rev. B \textbf{77}, 193309
(2008).

\bibitem{Pepino09} R. A. Pepino, J. Cooper, D. Z. Anderson, and M. J.
Holland, Phys. Rev. Lett. \textbf{103}, 140405 (2009).


\bibitem{Bhat06} R. Bhat, M. J. Holland, and L. D. Carr, Phys. Rev. Lett.
\textbf{96}, 060405 (2006).


\bibitem{Sagnac13} M. G. Sagnac, Compt. Rend. \textbf{157}, 708 (1913).

\bibitem{Post67} E. J. Post, Rev. Mod. Phys. \textbf{39}, 475 (1967).


\bibitem{Buttiker86} M. B\"{u}ttiker, Phys. Rev. Lett. \textbf{57}, 1761
(1986).

\bibitem{Azbel81} M. Y. Azbel, J. Phys. C \textbf{14}, L225 (1981).

\bibitem{Buttiker85} M. B\"{u}ttiker and Y. Imry, J. Phys. C \textbf{18},
L467 (1985).


\bibitem{Peierls33} R. Peierls, Physik Z. \textbf{80}, 763 (1933).


\bibitem{JinLiang09} L. Jin and Z. Song, arXiv: 0906.5049.

\bibitem{XiaJianBai92} J. B. Xia, Phys. Rev. B \textbf{45}, 3593 (1992).


\bibitem{Datta} S. Datta, \textit{Electronic Transport in Mesoscopic Systems}
(Cambridge University Press, Cambridge, 1995).


\bibitem{Amico05} L. Amico, A. Osterloh, and F. Cataliotti, Phys. Rev. Lett.
\textbf{95}, 063201 (2005).

\bibitem{Arnold07} S. F. Arnold, J. Leach, M. J. Padgett, V. E. Lembessis,
D. Ellinas, A. J. Wright, J. M. Girkin, P. \"{O}hberg and A. S. Arnold, Opt.
Express \textbf{15}, 8619 (2007).

\bibitem{Cramer} http://en.wikipedia.org/wiki/Cramer's\_rule.
\end{thebibliography}
\end{document}